\documentstyle[prb,aps,epsf,twocolumn]{revtex}
\begin{document}
\draft
\twocolumn[\hsize\textwidth\columnwidth\hsize\csname@twocolumnfalse%
\endcsname

\preprint{}

\title{Dual Vortex Theory of Strongly Interacting Electrons: 
A Non-Fermi Liquid with a Twist}
\author{Leon Balents}
\address{Room 1D-368, Bell Labs, Lucent Technologies, 700 Mountain Ave, Murray
  Hill, NJ 07974}
\author{Matthew P. A. Fisher}
\address{Institute for Theoretical Physics, University of
California, Santa Barbara, CA 93106-4030}
\author{Chetan Nayak}
\address{Physics Department, University of California, Los Angeles, CA 
  90095--1547}
\date{\today}
\maketitle

\begin{abstract}
As discovered in the quantum Hall effect, a very effective way for
strongly-repulsive electrons to minimize their potential energy is
to aquire non-zero relative angular momentum.  We pursue this mechanism 
for interacting two-dimensional electrons in zero magnetic field,
by employing a representation of the electrons as
composite bosons interacting with a Chern-Simons gauge field.
This enables us to construct a dual description in which the fundamental
constituents are vortices in the auxiliary boson fields.  The
resulting formalism embraces a cornucopia of possible phases.  Remarkably,
superconductivity is a generic feature, while the Fermi liquid is
not -- prompting us to conjecture that such a state may not be
possible when the interactions are sufficiently strong.  Many
aspects of our earlier discussions of the nodal liquid and
spin-charge separation find surprising incarnations in this new
framework.
\end{abstract}
\vspace{0.15cm}

]
\narrowtext

\section{Introduction}

Fermi liquid theory is
the cornerstone of the modern theory
of metals, as well as band theories of insulators
and semiconductors. This theory -- like
most perturbative theories -- is informed by the assumption
that the kinetic energy is the
dominant scale. As a result,
the theory is constructed in momentum space,
where the kinetic energy is diagonalized.
This leads to strong kinematic constraints
which circumscribe corrections to the
underlying free fermion behavior.
In this paper, our point of departure
is a different extreme limit in which
the interaction must be dealt with
at the outset, and the kinematic constraints might,
consequently, be inoperative. Hence,
we are forced to adopt a non-perturbative
approach. As is often the case in
non-perturbative problems -- for example 
the quantum Hall effect\cite{fqhe} and the one-dimensional
electron gas\cite{1deg} -- it is advantageous to adopt
a {\it real space} approach.  Here, this also enables us
to gain a vantage point from
which to focus on the strong electron interaction.

The past few decades
have witnessed the discovery of a number of
physical systems in which the interaction energy is
comparable to or greater than the kinetic energy.
These materials exhibit strange behavior
which is not readily captured within the conventional
Fermi liquid framework.
The high-$T_c$ cuprate superconductors\cite{cuprates}
are the most famous example, but there
are certainly others, such as heavy-fermion
materials\cite{heavy} and high-mobility
2DEGs at large $r_s$\cite{2d_mit}.
Ironically, even $^3$He -- the birthplace of
Fermi liquid theory -- falls in this category\cite{He3}.
The analysis of such systems may require
an approach of the type propounded
in this paper. 

Following the above reasoning, we are led to search for a means of
incorporating strong electron--electron repulsion from the outset.  As
discovered by Laughlin,\cite{Laughlin83}\ the spatial separation due
to the centrifugal barrier for non-zero angular momentum is a very
effective way for particles to lower their Coulomb energy.
We consider strongly interacting electrons moving
in the two-dimensional (2d) continuum,
and assume that these
strong interactions include a hard-core which prevents the crossing
of electron trajectories.  Some of the resulting
physics is reminiscent of the
quantum Hall effect: pairs of particles tend to spin around one
another.  There are, however, some significant differences:
time-reversal symmetry is not explicitly broken, and
further, the kinetic energy is not quenched.  Nevertheless,
our investigations in the remainder of this paper
and elsewhere\cite{unpub}\ lead us to suspect that
strongly repulsive electrons in the 2d continuum
can form a p-wave ($p_x \pm i p_y$) superconductor!  

To develop a low-energy effective field theory, we first use only the
non-crossing constraint on the fermion world-lines.  This is a
sufficient condition to allow the use of statistical
transmutation\cite{stat_trans,Fradkin}\ to realize up and down-spin
electrons as bosonic fields interacting with a Chern-Simons gauge
field which attaches flux to {\it spin}. {\sl Without additional
  assumptions}, we can then pass to a dual theory of vortices in the
up- and down-spin bosonic fields.  In this way, we argue that many of
our previous results on the nodal liquid\cite{BFN1,BFN2}\ hold with a
much wider range of validity.  With this approach, we believe that we
gain an unfettered view of the entire phase diagram of this {\it
  infinitely} strongly interacting fermionic system.

Since our dual theory is of Ginzburg-Landau form,
its phase structure can be analyzed by considering
the condensation of various fields. If no vortex field
condenses, the system is superconducting
with a non-zero angular momentum pairing
state with $p_x + ip_y$ symmetry\cite{laughlin}.
Since vortex condensation typically implies
charge ordering, it is usually driven
by a periodic potential or long-range Coulomb interactions.
In their absence
we thus conjecture that the generic state of the strongly-interacting
system is superconducting.
This is an astonishing conclusion,
given the lack of a palpable `pairing mechanism'.
Evidently, strongly repulsive interactions
essentially force electrons of opposite
spin to ``rotate" about one another and
introduce strong superconducting correlations.
Other ordered phases result when the vortex fields
condense.  For example, a spin- or charge-density wave results if  
both the up- and down-spin vortex fields condense.

A basic feature of {\it any} superconductor is {\it spin-charge}
separation\cite{kiv-rok}.  To access spin-charge separation in the
$p_x + i p_y$ superconductor within our dual Ginzburg-Landau
formulation {\it requires} consideration of ``paired" vortex
composites.  When these composite bosons condense, they can destroy
the superconductivity -- but spin-charge separation survives.
Specifically, if a vortex in the down-spin boson field pairs with a
vortex in the up-spin boson field and this pair condenses,
translational symmetry is spontaneously broken by the formation of a
crystalline state of {\it spinless} charge $e$ solitons.  The spin
sector is gapped except for chiral edge states, so that this phase is
a $T$-violating nodal liquid (i.e. a chiral spin liquid).
Alternatively, if a vortex in the down-spin boson field pairs with an
{\it anti}-vortex in the up-spin boson field and this pair condenses,
a transition occurs into a fully-gapped superconductor, such as a
superconductor with tightly-bound pairs.  If both types of vortex
pairs condense, an analogous spin-liquid results.  In each of these
phases, time reversal invariance is spontaneously broken.

With the inclusion of an ionic potential
acting on the electrons, however, $T$-invariant phases
are possible, and expected.  Indeed, by allowing 
for terms in the dual Ginzburg-Landau theory
which break rotational invariance,
gapped modes in the spin sector can go
soft at finite momentum.  For a uniaxial potential, gapless
modes naturally appear at two
points in momentum space.
At these two points
the vortex-anti-vortex field
is critical, and can
be conveniently re-fermionized as two Dirac fields.
These can be identified as the 
nodal quasiparticles
of a $p_x$ superconducting phase.
Similarly, an ionic potential with square symmetry leads to four
low-energy points in momentum space, and thereby a fourfold Dirac theory
recovering the spectrum of $d_{x^2-y^2}$ quasiparticles (see below and 
Sec.~VI for a discussion of some subtleties of the d-wave case). 
Within our theory, a 
(very) strong local repulsion acting in concert
with an
ionic potential with square symmetry are
the essential ingredients for high
temperature $d$-wave superconductivity.
In the {\it absence} of the ionic potential,
strong  $T$-violating pairing with $p_x + i p_y$ symmetry
is expected.  This is the pairing symmetry in
the A-phase of a superfluid 3-He
film\cite{Vollhardt}.

An appealing feature of our dual Ginzburg-Landau
formulation
is that it gives a clear meaning to
spin-charge separation (and spin-charge confinement)
in two-dimensional electron systems.
Indeed, we identify an Ising-like $Z_2$ symmetry
which when unbroken leaves spin and charge separate.
Spin-charge confinement is
driven by an Ising ordering transition.

Remarkably, although our theory is intimately tied to a real-space
picture, Fermi surface physics is {\it not} lost, as evidenced by the
nodal quasiparticles.  The {\sl Fermi liquid} phase itself is much
more elusive!  It occurs, if at all, only as a narrow region analogous
to a (fermionic) ``entangled flux liquid'' in the Ginzburg-Landau theory, and
is certainly only possible if the $Z_2$ symmetry is broken.
However, it is unlikely that such a state can be
obtained in the Ginzburg-Landau theory.
It is here that we speculate the relative angular momentum
between electrons plays an
important role.  Indeed, (very singular) s-wave scattering states are
known to dominate the physics of the Fermi liquid with hard-core
radius $a \ll 1/k_F$.  We suspect that the inclusion of such singular
wavefunctions invalidates the passage to a {\sl local and analytic}
vortex lagrangian.  See Sec.~VI for further discussion of this point.

The phase diagram which results from this analysis contains a plethora
of fascinating states, including superconducting states of pairing
symmetry $p_x$, ${p_x} + i {p_y}$, $d_{{x^2}-{y^2}}$, $d_{xy}$, and
their quantum disordered counterparts.  These states are characterized
by a separation between the characteristic scales of the charge and
spin and even -- when the $Z_2$ symmetry is unbroken -- true
spin-charge separation.  A particularly salient aspect of this phase
diagram is the miniscule domain, and we speculate perhaps even
absence, of the Fermi liquid state.  This seems to support the claims
of Anderson\cite{pwa_nfl}\ and co-workers that
Fermi liquid theory breaks down in
strongly-interacting two-dimensional systems (though it does not
support their bold claims that Fermi liquid theory breaks down even at
weak-coupling). The proofs of the stability of Fermi liquid theory at
weak-coupling\cite{flt}\  presumably do not apply to the strongly-interacting
limit which we have considered.

In section II, we first discuss the statistical transmutation which
obviates the need for a local pair field. Constructing
a dual theory, we describe the standard antiferromagnetic
and charge-density-wave states which result from
the condensation of single vortices. In section III,
we discuss paired vortices and the $Z_2$ symmetry
which distinguishes their condensation
from that of individual vortices. When this symmetry is
unbroken, spin and charge separate. In section IV,
we discuss the phase diagram which results from the
condensation of paired vortices. This phase
diagram revolves about a $p_x\pm i p_y$ superconducting
state. In section V, we show how time-reversal (T) invariant
superconducting states
such as $p_x$ or $d_{xy}$ 
can arise in this model. The momentum-space
structure and concomitant phenomenology
of BCS-like $d_{xy}$ or $d_{x^2-y^2}$ superconductors
is recovered.  We find that the physics of the nodal liquid
reappears in a new guise: the nodons are vortices
in a vortex field whose fermionic statistics
result from their interaction with a Chern-Simons
gauge field. In chapter VI, we arrive, ultimately,
at a phase diagram which is the synthesis
of ideas of duality and vortex condensation
common to field theories of the quantum Hall effect
as well as our earlier work on nodal liquids,
but is almost entirely orthogonal to the underlying
conceit of Fermi liquid theory.

\section{Fermions, Flux Attachment, and Duality}

We focus throughout on spinful electrons moving in
the two-dimensional continuum, interacting
via a spin-independnet interaction.
We assume that the electron-electron repulsion
is strong enough that no two electrons can ever be coincident.
Precisely this ``hard-core'' constraint makes  it possible to 
transform the interacting two-dimensional electron gas
into a mathematically equivalent system of
interacting spinful {\it bosons},
by attaching ``statistical" flux with
an appropriate
Chern-Simons gauge field.  Such a ``bosonization"
scheme for 2d {\it spinless} electrons
has been particularly illuminating in
the context of the fractional quantum Hall effect\cite{fqhe,Fradkin}.
With spin there is considerable freedom
in how one attaches the flux tubes to convert fermions
into bosons.  We adopt a scheme in which
flux is attached to the {\it spin}
of the electrons, and define  
\begin{equation}
c_\alpha(\bbox{r}) = b_\alpha(\bbox{r}) \exp[ie_\alpha \int_{\bbox{r}'}
\Theta(\bbox{r}-\bbox{r}')2S^z(\bbox{r}')] ,
\label{eq:CStransform}
\end{equation}
with a ``charge" $e_\uparrow = 1$ and $e_\downarrow = -1$.
Here \begin{equation}
S^z(\bbox{r}) = [n_\uparrow(\bbox{r}) - n_\downarrow(\bbox{r})]/2
\end{equation}
is the z-component of the spin density operator
with $n_\alpha = c^\dagger_\alpha c_\alpha^{\vphantom\dagger} = b^\dagger_\alpha
 b_\alpha^{\vphantom\dagger}$
(no sum on $\alpha$), and $\Theta(\bbox{r})$ denotes
the angle that $\bbox{r}$ makes with the $x$-axis.
The boson operators satisfy
canonical commutators, $[b_\alpha^{\vphantom\dagger}({\bbox{r}}),
b_\beta^\dagger({\bbox{r}'})]
=0$ for $\bbox{r} \ne \bbox{r}'$.  Due to the 
non-crossing constraint, the ``onsite" commutators
need not be specified.  

An advantage
of the
above scheme for flux attachment is that with
zero total spin, $S^z_{tot} = 0$ (as assumed hereafter),
the statistical flux ``seen" by the Chern-Simons bosons
{\it vanishes} on average.  For spinless
electrons this happy situation
requires the presence of a strong external magnetic field
(as in the FQHE, c.f. Ref.~\onlinecite{fqhe}).
As for Abelian bosonization
in one spatial dimension\cite{Fradkin},
the above choice of a spin quantization axis
masks the underlying spin-rotational invariance.
But as we shall see, it is possible
to restore explicit SU(2) symmetry by a subsequent
``refermionization".  

After transforming to boson operators,
the partition function for the 
Hamiltonian of 2d interacting electrons can be expressed
as a functional integral over bosonic fields
and a statistical gauge field, $\alpha_\mu$,
with associated (Euclidean) Lagrangian density:
\begin{equation}
{\cal L} =  b_\alpha^* (\partial_\tau - ie_\alpha \alpha^0)
b_\alpha^{\vphantom\dagger}
-{1 \over {2m_e}} b_\alpha^* (\nabla - ie_\alpha \bbox{\alpha} )^2
b_\alpha^{\vphantom\dagger}
- {\cal L}_{cs}  .
\label{eq:composite_boson}
\end{equation}
In the ``Coulomb gauge" $\partial_i \alpha^i =0$
the Chern-Simons term is ${\cal L}_{cs} = (i/2\pi) \alpha^0
(\epsilon_{ij}\partial_i\alpha_j)$,
but can be cast into a more convenient
gauge invariant form:
\begin{equation}
{\cal L}_{cs}(\alpha_\mu) = i {1 \over {4\pi}} \epsilon_{\mu \nu \lambda}
\alpha_\mu \partial_\nu \alpha_\lambda   .
\end{equation}
The form of the electron interaction term (not shown) is unchanged
under ``bosonization" due to the
equivalence of the fermion and boson densities:
$c^\dagger c = b^\dagger b$.

We now implement the standard
2+1-dimensional duality transformation\cite{ftduality}
which exchanges bosons (the $b_\alpha$'s)
for vortices in the bosonic fields - arriving at a description
in terms of 
vortex field operators, denoted $\Phi_\alpha$. 
To illuminate this, it is instructive
to briefly consider an alternate representation
in terms of boson world lines:
\begin{equation}
{\cal L}_{wl}  = {1 \over 2} (J_\mu^\alpha)^2 + i e_\alpha 
J^\alpha_\mu \alpha_\mu  - {\cal L}_{cs}(\alpha_\mu)  .
\end{equation}
Here $J^\alpha_\mu$ denotes a bosonic three-current 
(with $\mu$ running over 2+1 space-time coordinates) for
spin component $\alpha$.  The first term measures
the length of the space-time world-lines and represents the kinetic
energy.  The Chern-Simons coupling generates
a sign change when two world lines exchange,
transforming to fermions.  To implement
duality, these three-currents are expressed
in terms of two gauge fields, $a^\alpha_\mu$, one for each
spin component:
\begin{equation}
J^\alpha_\mu = {1 \over {2\pi}} \epsilon_{\mu \nu \lambda} \partial_\nu a^\alpha
_\lambda  .
\end{equation}
In this way, 
charge conservation
($\partial_\mu J^\alpha_\mu =0$) is automatically satisfied.
The dual Lagrangian can be obtained
by inserting this expression into ${\cal L}_{wl}$,
and integrating out the Chern-Simons field, $\alpha^\mu$.  
Electron charge {\it quantization} is implemented
by the vortex operators, $\Phi_\alpha$,
which are minimally coupled to $a^\alpha_\mu$.  

The
final dual form consists of two Ginzburg-Landau theories,
coupled via a Chern-Simons term:
\begin{equation}
{\cal L}_{dual} = \sum_\alpha {\cal L}_{GL}(\Phi_\alpha,a^\alpha_\mu) + {\cal L}
_{cs}(a^\uparrow_\mu - a^\downarrow_\mu)    ,
\end{equation}
\begin{equation}
{\cal L}_{GL}(\Phi, a_\mu) = 
 {1 \over 2} |(\partial_\mu - i a_\mu)\Phi |^2
+ V(\Phi) + {1 \over 2} (f_{\mu \nu})^2   .
\label{eqn:GL_Lag}
\end{equation}
with a  ``potential" that can be expanded as
$V(\Phi) = r|\Phi|^2 + u|\Phi|^4+...$.  

Since the Chern-Simons term
only involves {\it spin} currents, it is extremely
convenient to introduce charge and spin gauge fields:
\begin{equation}
a_\mu^{\rho} = a^\uparrow_\mu + a^\downarrow_\mu    ;  \hskip0.5cm
a_\mu^{\sigma} = a^\uparrow_\mu - a^\downarrow_\mu     ,
\end{equation}
(and corresponding field strengths, $f_{\mu \nu}^{\rho},f_{\mu \nu}^\sigma$).
As with Abelian bosonization in one dimension,
charge and spin currents defined by $J_\mu^\rho = J_\mu^\uparrow + J_\mu^\downarrow$ and $J_\mu^\rho = J_\mu^\uparrow - J_\mu^\downarrow$,
are given by derivatives
of the charge and spin fields, respectively:
\begin{equation}
J^\rho_\mu = {1 \over {2\pi}} \epsilon_{\mu \nu \lambda} \partial_\nu a^\rho
_\lambda  ; \hskip0.5cm    J^\sigma_\mu = {1 \over {2\pi}}
\epsilon_{\mu \nu \lambda} \partial_\nu a^\sigma_\lambda  
\label{current}.
\end{equation}
Longer range Coulomb interactions
can be readily incorporated by adding a term
bilinear in the electron charge density: $\epsilon_{ij} \partial_i
a^\rho_j$.
In this dual representation,
$\Phi_\alpha^\dagger$ creates 
a vortex in the {\it electron} wavefunction - effectively
increasing the angular momentum of {\it all} spin $\alpha$
electrons by one unit.  An important feature of ${\cal L}_{dual}$
is that there are precisely as many positive as negative
circulation (electron) vortices (for each spin species),
implying a relativistic form for $\Phi_\alpha$.

It is instructive to briefly mention how
these dual fields couple to an external electromagnetic
field, $A_\mu$.  As usual $A_\mu$ couples directly
to the total electrical current,
$J_\mu^\rho$,
so that from Eq.~\ref{current}  one has:
\begin{equation}
{\cal L}_A = {1 \over {2 \pi}} A_\mu 
\epsilon_{\mu \nu \lambda} \partial_\nu a_\lambda^\rho   .
\end{equation}
It is also convenient to introduce an external ``spin" gauge field,
$A_\mu^\sigma$, which couples to the total
(z-component) spin current, $J_\mu^\sigma$:
\begin{equation}
{\cal L}_A^\sigma = {1 \over {2 \pi}} A_\mu^\sigma 
\epsilon_{\mu \nu \lambda} \partial_\nu a_\lambda^\sigma   .
\label{spin-source}
\end{equation}

The dual Ginzburg-Landau representation can be fruitfully employed
to describe various possible phases of spinful 2d electrons
satisfying the `non-crossing' constraint.
For instance, imagine a phase in which
the (electron) vortices are absent in the ground state
(except as virtual fluctuations),
which corresponds
to taking $r$ large and positive in the
above Ginzburg-Landau description.  Being massive, $\Phi_\alpha$
can be safely integrated out, leaving
an effective theory:
\begin{equation}
  {\cal L}_{eff} = {1 \over 2} (f^\rho_{\mu \nu})^2 +
  {\cal L}_{cs}(a^\sigma_\mu) ,
  \label{eq:LeffCS}
\end{equation} 
which describes massless charge fluctuation, and spin
fluctuations gapped (in the bulk) by the Chern-Simons term.
This is a superconducting phase, which can be verified
by noting that the pair field operator
($c_\uparrow c_\downarrow$) creates
a (2+1 space-time) monopole of strength two in the
field $B^\rho_\mu = \epsilon_{\mu \nu \lambda} \partial_\nu
a^\rho_\lambda $.  Since the gauge field $a^\rho_\mu$
is massless, the energy cost
to make a free monopole is finite (monopoles
interact via 2+1 Coulomb force), so that
the pair-field exhibits true ODLRO.
Due to the Chern-Simons term, this superconducting
phase exhibits a quantized Hall ``spin-conductance",
$\sigma^s_{xy} = 1$, a signature of 
a spontaneous breakdown of time reversal invariance.
This follows by noting that the Lagrangian with spin ``gauge" field,
${\cal L}_{eff} + {\cal L}_A^\sigma$, depends quadratically
on both $a^\rho_\mu$ and $a^\sigma_\mu$, so that they can be integrated
out to give,
\begin{equation}
{\cal L}_{eff} + {\cal L}_A^\sigma = -i \sigma_{xy}^s {1 \over {4\pi}} \epsilon_
{\mu \nu \lambda}
A^\sigma_\mu \partial_\nu A^\sigma_\lambda   ,
\end{equation}
with $\sigma_{xy}^s =1$.
Following the analysis in Ref.~\onlinecite{SF98}, one can
readily verify that a 2d BCS superconductor
with $p_x + i p_y$ pairing symmetry has precisely
such a value for the quantized spin conductance (also
see below in Section IV).
The spin state of the pair is then presumably
a triplet with $s_z = 0$.  This is the phase
of a 2d superfluid 3-He A film.    

Before discussing spin-charge
separation, which is a generic property of a 2d superconductor,
it is instructive to consider phases described
in the dual theory when vortices created by $\Phi_\alpha$
proliferate, and condense: $\langle \Phi_\alpha \rangle \ne 0$.
As we shall see, in contrast to the $p_x + ip_y$ superconducting
phase, these phases typically exhibit crystalline
order, spontaneously breaking {\it translational}
symmetry.  To see this, note that 
upon vortex condensation, the dual ``flux" $\epsilon_{ij} 
\partial_i a^\alpha_j$ is quantized in units of $2\pi$, which corresponds
to quantization of charge in units
of the electron charge $e$.  By analogy with
the Abrikosov flux lattice phase of a Type II superconductor, one expects
a breakdown of translational symmetry
with spin up (and down) electrons forming an ordered
lattice.  Depending on the relative phase
between the density wave of spin up and down electrons,
this will be either a charge density wave (CDW) state
or an antiferromagnet (AF).  In the presence
of a commensurate background periodic potential from the ions
in the solid, one expects these density wave states
to lock, resulting in insulating behavior.

These ``crystalline" phases can presumably be energetically stabilized
by a longer-range repulsive interaction between the electrons, in
addition to the `non-crossing' constraint (which is required to make
our 2d bosonization scheme legitimate).  In their absence, our dual
Ginzburg-Landau representation, (fortified by subsequent analysis
below and physical reasoning) strongly suggests that the predominant
ground state is the $p_x+ip_y$ (or $p_x -ip_y$) superconductor.  Given
that spin-up and spin-down electrons prefer a state of non-vanishing
relative angular momentum to minimize Coulomb repulsion, an $\ell =
\pm 1$ orbital angular momentum state is clearly favored by the
kinetic energy.  As such, it seems that incorporating local Coulomb
repulsion by forcing electrons of opposite spin into a relative
angular momentum state is a {\it very} effective electronic mechanism
for high temperature superconductivity!

\section{Spin-Charge Separation}

The phases described above are not
the only possible phases for 2d electrons
satisfying the `non-crossing' constraint.
Composite order parameters can also condense,
thereby leading to charge- and/or spin-insulators.
We focus on the combinations
\begin{equation}
{\Phi_\rho}={\Phi_\uparrow}{\Phi_\downarrow}   ;  \hskip0.7cm
{\Phi_\sigma}={\Phi_\uparrow}{\Phi^\dagger_\downarrow}  ,
\end{equation}
which, as we shall see, are exceedingly interesting from a
phenomenological standpoint.  These order parameters can condense
without breaking the $Z_2$ symmetry
\begin{equation}
\label{eqn:z_2}
{\Phi_{\uparrow,\downarrow}}\rightarrow
- {\Phi_{\uparrow,\downarrow}}  ,
\end{equation}
so we can have
$\langle {\Phi_{\rho,\sigma}}\rangle\neq 0$ while
${\Phi_{\uparrow,\downarrow}}=0$. 

In the following we presume that $\Phi_\rho$ and $\Phi_\sigma$
describe the soft modes at low energies, and that
$\Phi_\alpha$ remains massive.  
As we shall see, this leads naturally
to a separation of low energy spin and charge degrees of freedom.
Our motivation for this is two-fold.
Firstly, spin-charge separation is a generic property
of a superconductor such as the $p_x + ip_y$
state discussed above,
and it is instructive to
exhibit this separation within the present
Ginzburg-Landau framework.  But secondly,
in many Mott insulators of interest
the charge degrees of freedom freeze out at
much higher energy scales than the
energies on which local moments and spin
order develops.  This is typified
by the undoped cuprate materials,
with insulating behavior setting in
on the scale of electron volts (the ``Hubbard" $U$)
much higher than the antiferromagnetic ordering temperature.
In order to capture these two very different
energy scales within the present framework, it is essential
to transform to the charge and spin vortex fields,
$\Phi_\rho$ and $\Phi_\sigma$.  Indeed,
in the description of the antiferromagnetic insulator
discussed above driven by condensation of
$\Phi_\alpha$, charge ordering and local moment
formation
necessarily take place on the {\it same} energy scale,
since the dual flux tubes in these vortex fields
are electrons carrying {\it both} charge and spin.

Under the assumption that both fields $\Phi_\alpha$
remain massive, one can write down an effective theory
for the soft modes ${\Phi_{\rho,\sigma}}$ by
integrating out ${\Phi_{\uparrow,\downarrow}}$.
Below we illustrate how this can be done, by
regularizing the theory on a lattice.  But more generally,
the form of the effective theory is essentially dictated
by symmetries, involving three contributions: 
\begin{equation}
{\cal L}_{eff} = {\cal L}_\rho + {\cal L}_\sigma + {\cal L}_{int}  ,
\label{eq:eff_Lag}
\end{equation}
with a charge sector,
\begin{eqnarray}
{\cal L}_\rho  &=& 
 {1 \over 2} {|(\partial_\mu - i{a^\rho_\mu}){\Phi_\rho}|^2}
+ {r_\rho}{|{\Phi_\rho}|^2} + u_\rho |\Phi_\rho|^4 \cr
& &\,\,\,\,+ {1 \over 2} {(f^\rho_{\mu \nu})^2}
+ {1 \over {2\pi}} A_\mu {\epsilon_{\mu \nu \lambda}} 
\partial_\nu {a^\rho_\lambda}   ,
\end{eqnarray}
a spin sector,
\begin{eqnarray}
{\cal L}_\sigma = &=& {1 \over 2} {|(\partial_\mu - i{a^\sigma_\mu}){\Phi_\sigma
}|^2}
+ {r_\sigma}{|{\Phi_\sigma}|^2} + u_\sigma |\Phi_\sigma|^4 \cr
& &+ 
{1 \over 2} {(f^\sigma_{\mu \nu})^2}
+ i {1 \over {4\pi}} \epsilon_{\mu \nu \lambda}
{a^\sigma_\mu} {\partial_\nu} {a^\sigma_\lambda}
\label{L-spin}
\end{eqnarray}
and sub-dominant interaction terms involving
many derivatives (see eg. below).
The charge sector has the Ginzburg-Landau form,
with minimal coupling to the charge gauge field,
$a^\rho_\mu$, and $A_\mu$ is the physical
electromagnetic potential.  The Chern-Simons term
lives solely in the spin sector.

Some insight into the genesis of such
a Lagrangian may be obtained by considering a lattice
version of (\ref{eqn:GL_Lag}) and dropping
the $a^{\uparrow,\downarrow}_\mu$ for simplicity.
Writing ${\Phi_\alpha}={e^{i\theta_\alpha}}$, we have
\begin{equation}
S = {\sum_{\langle i,j\rangle}}\cos({\theta_i^\alpha}-{\theta_j^\alpha})  ,
\end{equation}
where $i$ and $j$ denote sites
of a 2d (say square) lattice and a sum over $\alpha$ is understood.
We now introduce charge and spin fields ${\theta^{\rho,\sigma}}$:
\begin{equation}
{\theta^{\uparrow,\downarrow}}= \frac{1}{2}({\theta^{\rho}}\pm
{\theta^{\sigma}}) + \frac{\pi}{2}s
\end{equation}
where $s=\pm 1$ is an Ising ``spin" variable. By introducing $s$, we can treat
${\theta^{\rho,\sigma}}$ as angular variables
since the action is invariant under
${\theta^{\rho,\sigma}}\rightarrow {\theta^{\rho,\sigma}}+2\pi$,
$s\rightarrow -s$.
The action can then be rewritten as:
\begin{equation}
S = {\sum_{\langle i,j\rangle}} {s_i}{s_j}\,
\cos\frac{1}{2}({\theta_i^\rho}-{\theta_i^\rho})
\cos\frac{1}{2}({\theta_i^\sigma}-{\theta_i^\sigma}) ,
\end{equation}
since $\sin{\pi\over 2}({s_i}-{s_j}) = 0$
and $\cos{\pi\over 2}({s_i}-{s_j})={s_i}{s_j}$.
Let us now consider the effect of integrating out
the ${s_i}$'s. If we are in the symmetric phase
in which the $Z_2$ is unbroken, this can be done
perturbatively, as in the high-temperature expansion
for the Ising model.
\begin{eqnarray}
Z &=& {\sum_{{s_i}=\pm 1}}{e^{\sum_{ij} \beta J {s_i}{s_j}}}\cr
&=& {\sum_{{S_i}=\pm 1}}{\sum_n}\frac{1}{n!}
{\left({\sum_{ij}\beta J {s_i}{s_j}}\right)^n}\cr  
\end{eqnarray}
To leading order in $\beta$, which corresponds to decoupled
free spins, one has $\langle s_i s_j \rangle = \delta_{ij}$,
which implies an effective action of the form:
\begin{eqnarray}
S_{eff} &=& {\sum_{\langle i,j\rangle}} \,
\left(1+\cos({\theta_i^\rho}-{\theta_j^\rho})\right)
\left(1+\cos({\theta_i^\sigma}-{\theta_i^\sigma})\right)\cr
 &=& {\sum_{\langle i,j\rangle}} \,
\cos({\theta_i^\rho}-{\theta_j^\rho})+
\cos({\theta_i^\sigma}-{\theta_j^\sigma})\cr
& &+ \cos({\theta_i^\rho}-{\theta_j^\rho})
\cos({\theta_i^\sigma}-{\theta_j^\sigma}) .
\end{eqnarray}
Upon making the identifications
${\Phi_\rho}={e^{i\theta_\rho}}$, ${\Phi_\sigma}={e^{i\theta_\sigma}}$
and restoring the gauge fields (minimally coupled)
the first two terms are seen to be lattice versions
of the continuum Ginzburg-Landau theories
in ${\cal L}_\rho$ and ${\cal L}_\sigma$, respectively.
The last term generates
a gradient interaction term
between the charge and spin sectors.

When the $Z_2$ symmetry is unbroken, as it is by assumption
in (\ref{eq:eff_Lag}), spin and charge separate, as
we now argue. In the $p_x + ip_y$ superconducting state, which can
be described by either (\ref{eqn:GL_Lag})
with ${r_{\uparrow,\downarrow}}>0$
or (\ref{eq:eff_Lag}) with ${r_{\rho,\sigma}}>0$,
the low-energy excitations are the gapless
superfluid mode, ${a^\rho_\mu}$, which carries
charge but no spin. At finite energy, there
are also the quanta of $\Phi_\sigma$, which are fermionic and
carry spin-$1/2$ by virtue of their coupling to the
Chern-Simons gauge field, ${a^\sigma_\mu}$.
They do not couple directly to the electromagnetic field,
so we assign them quantum numbers $q=0, s=1/2$.
As we shall see in the next section,
these neutral fermionic spin 1/2 excitations
are the p-wave analog of 
``nodons", introduced in reference \onlinecite{BFN1}
for a d-wave superconductor.

When ${r_{\rho}}<0$ and $\Phi_\rho$ condenses, the dual flux
$\epsilon_{ij} \partial_i a_j^\rho$ becomes quantized into
``flux-tubes", by direct analogy with a Type II Ginzburg-Landau
superconductor.  Each one of these dual `flux tubes" carries one unit
of electric charge, but no spin.  We refer to these spinless charge
$e$ solitons as ``holons"\cite{holons}.  We thus see that
provided the $Z_2$ symmetry in Eq. \ref{eqn:z_2} is {\it unbroken}, spin and
charge are separated.  On the other hand, when the $Z_2$ symmetry is
broken by the condensation of ${\Phi_{\uparrow,\downarrow}}$, the spin
and charge are confined.  Since ${\Phi_{\uparrow,\downarrow}}$ couples
to ${a^\rho_\mu}\pm{a^\sigma_\mu}$, this condensation locks the spin
and charge together, leaving only the electron in the spectrum.

To summarize, states 
of higher symmetry have less restricted spectra.
The original dual representation in Eq. \ref{eqn:GL_Lag} has a 
$U(1)\times U(1)$ gauge symmetry, corresponding
to independent rotations of $\Phi_\uparrow$
and $\Phi_\downarrow$; 
\begin{equation}
\Phi_\alpha \rightarrow \Phi_\alpha e^{i\Lambda_\alpha}   ;  \hskip0.7cm
a^\alpha_\mu \rightarrow a^\alpha_\mu + \partial_\mu \Lambda_\alpha  ,
\end{equation}
with two arbitrary functions $\Lambda_\alpha$.  This gauge symmetry
emerges when the conserved electron three-currents are expressed as a
curl of the gauge fields, $a^\alpha_\mu$.  Breaking down this large
symmetry corresponds to ``localization" or ``quantization" of charge
and/or spin.  When the full symmetry is completely broken, both charge
and spin become quantized together, and all the excitations have
quantum numbers of the electron, with ${q \over 2} + s$ an integer (as
in the antiferromagnetic insulator mentioned in Section II).  But if
this symmetry is only {\it partially} broken by condensation of $\Phi_\rho$
and $\Phi_\sigma$, leaving an unbroken $Z_2$, both charge ($e$) and
spin ($1/2$) become quantized, but excitations exist with any
combination of these quantum numbers.  This will be nicely illustrated
in the next Section where we employ the dual description,
Eq.~\ref{eq:eff_Lag}, to examine the properties of
some T-breaking spin-charge separated phases.

\section{Phases with Broken T}
 
Having established the form of Eq.~\ref{eq:eff_Lag} under
the assumption of an unbroken $Z_2$ symmetry,
and the concomitant spin-charge separation, we explore
possible phases which emerge from this effective
theory.  We first focus
on phases which (spontaneously) break
time reversal invariance.  As we shall see, these
emerge naturally for electrons
in jellium, moving in the absence of ionic
potentials.  In the following sections
we consider the effects of ionic
potentials which break rotational invariance
and naturally drive transitions
into time reversal invariant phases.

As already discussed, the phase in the absence of
either spin or charge vortices is a $p_x + ip_y$ superconductor.
Since the charge and spin sectors have effectively
decoupled under the assumption of
the unbroken $Z_2$ symmetry, it is possible to
consider them separately.
If the vortices in the charge sector proliferate and condense, 
$\langle \Phi_\rho \rangle \ne 0$, 
the field $\epsilon_{ij} \partial_i a^\rho_j$
becomes quantized in ``dual" flux tubes,
as discussed above.  Each of these ``flux" tubes 
carries charge $e$, but no spin.
These charge $e$ spinless ``holons"
are expected to crystallize,
by direct analogy with the Abrikosov flux-lattice.
This charge $e$ crystal will presumably lock
to any underlying ionic potential.
With gapless spin-carrying edge states
still present, this electrically insulating phase is the $p_x+ip_y$ analog
of the nodal liquid.  Once again, energetic stabilization of this
crystalline phase presumably requires the presence of appreciable
Coulomb repulsion between electrons
on the scale of the mean electron spacing
(in addition to the `non-crossing' constraint).  

But suppose the spin vortices
condense, in the absence of charge vortices?
Due to the dual Anderson-Higgs mechanism, $a^\sigma_\mu$
becomes massive rendering the Chern-Simons term ineffective,
and leading to a 
spin-gap both in the bulk {\it and} at the edge.
This implies ``spin-insulating" behavior
with $\sigma^s_{xy} =0$.
How can we understand this fully spin-gapped
superconducting phase?  To this end
it is convenient to briefly consider
a BCS description of the quasiparticles in a $p_x +ip_y$ superconductor:
\begin{equation}
{\cal H}_{BCS} = \sum_{\bbox{k}} \epsilon_{\bbox{k}} 
c^\dagger_{\bbox{k}\alpha} 
c_{\bbox{k}\alpha}^{\vphantom\dagger} + \Delta_{\bbox{k}} c^\dagger_{\bbox{k} \uparrow} c^\dagger_{
-\bbox{k} 
\downarrow} + h.c. ,
\label{eq:BCSham}
\end{equation}
with dispersion $\epsilon_{\bbox{k}}= (k^2/2m_e) - \mu$
and gap function $\Delta_{\bbox{k}} = v_\Delta (k_x + i k_y)$.
In terms of a two-component spinor,
$\psi_1({\bbox{r}}) = c_\uparrow({\bbox{r}})$
and $\psi_2({\bbox{r}}) = c^\dagger_\downarrow({\bbox{r}})$,
this can be rewritten in the form of a Dirac equation
with (Euclidean) Lagrangian:
\begin{equation}
{\cal L}_{BCS} =  \psi^\dagger [\partial_\tau + \tau^z
((-\partial_j^2/2m_e) - \mu)
+ i v_\Delta \tau^j \partial_j  ] \psi  ,
\label{BCS-Dirac}
\end{equation}
with $j=x,y$.  This gives the usual 
BCS quasiparticle dispersion:  
$E_{\bbox{k}} = \pm \sqrt{\epsilon_{\bbox{k}}^2 
+ 
(v_\Delta k )^2 }$.
As in Ref.~\onlinecite{BFN1}, one can define a gauge invariant
charge neutral quasiparticle (a ``nodon"), by transforming
$\psi \rightarrow \exp(i\tau^z \varphi/2) \psi$,
with $\varphi$ the phase of the complex pair-field.
Spin and charge are thereby separated,
with the z-component of spin being the conserved
U(1) ``charge" in the Dirac theory:  $2S^z = \psi^\dagger \psi$.
Since the source field $A_\mu^\sigma$ couples
to the conserved spin current, it can be readily incorporated
into the above Dirac equation via a ``minimal coupling"
prescription: $\partial_\mu \rightarrow \partial_\mu - i A_\mu^\sigma$.  
  
In the presence of a boundary, say at
$y=0$ with 
boundary conditions $c_\alpha(x,y=0)=0$, one can readily show from
the above Dirac theory that a
chiral fermion edge mode exists only for positive
chemical potential, $\mu >0$.  In this BCS limit
one clearly has $\sigma^s_{xy} = 1$.  But at very strong coupling
when $\mu$ changes sign, the ground state
changes to a paired ``molecular"\cite{molecular}
limit with zero $\sigma^s_{xy}=0$.
Right at the transition, there are
gapless bulk quasiparticle excitations
described by a {\it massless} Dirac theory (at $\mu=0$ in Eq.~\ref{BCS-Dirac})
with a ``node" at zero momentum.  To access the molecular
limit of the $p_x+ip_y$ superconductor
presumably requires a very strong (and unphysical) {\it attractive}
interaction between electrons, enabling up and down spin electrons
to form a finite angular momentum {\it bound} state (with $\ell =1$).
The attractive interaction must overcome
the centrifugal {\it repulsion} between the two electrons
(present due to the `non-crossing' constraint).  

A direct connection between the molecular $p_x+ip_y$
superconductor and the phase
described by the dual theory
when the spin vortex condenses,
$\langle \Phi_\sigma \rangle \ne 0$, can be established by
re-fermionizing the spin sector of the Ginzburg-Landau theory
and showing its equivalence to the Dirac theory Eq.~\ref{BCS-Dirac}.
To illustrate this we instead
{\it bosonize} the Dirac theory.
In the BCS limit of the $p_x + i p_y$
superconductor with $\mu$ positive, the
massive relativistic Dirac fermion, $\psi$,
can be converted to a relativistic boson, $\Phi$,
via a Chern-Simons transformation:
\begin{eqnarray}
{\cal L}_{BCS} &=& {1 \over 2} |(\partial_\mu -i a_\mu -iA^\sigma_\mu )\Phi|^2
+ M^2 |\Phi|^2 + U |\Phi|^4 \cr
& &+ {\cal L}_{cs}(a_\mu) 
- {\cal L}_{cs}(A^\sigma_\mu). 
\label{L-BCS}
\end{eqnarray}
Here $M >0$ can be equated with the Dirac mass $\mu$.  Indeed,
the spectrum of this massive boson field, $\Phi$, is
$\omega_{\bbox{k}} = \pm \sqrt{M^2 + k^2}$ - 
the same form
as the BCS quasiparticle dispersion, $E_{\bbox{k}}$.
Since $\Phi$ is massive 
it can be safely integrated out.
The only remaining dependence on the source
field $A^\sigma_\mu$ is through the last
Chern-Simons term, which has been included
to give the correct
result for the spin Hall conductivity:  $\sigma^s_{xy}=1$.

Remarkably, Eq.~\ref{L-BCS} {\it precisely} coincides
with the {\it spin} sector
of the dual Ginzburg-Landau theory, ${\cal L}_\sigma$
in Eq.~\ref{L-spin}.
Indeed, with inclusion of the source term, ${\cal L}_A^\sigma$
in Eq.~\ref{spin-source}, the full Lagrangian
in the spin sector, ${\cal L}(A^\sigma_\mu) = {\cal L}_\sigma
+ {\cal L}_A^\sigma$ can be conveniently rewritten
by shifting $a_\mu^\sigma \rightarrow a_\mu^\sigma + A_\mu^\sigma$ as,
\begin{eqnarray}
  {\cal L}_\sigma(A_\mu^\sigma) &=& {1 \over 2} {|(\partial_\mu -
    i{a^\sigma_\mu}-iA_\mu^\sigma){\Phi_\sigma }|^2}
  +{r_\sigma}{|{\Phi_\sigma}|^2} + u_\sigma |\Phi_\sigma|^4 \cr
  & &+ 
  {\cal L}_{cs}(a_\mu) - {\cal L}_{cs}(A^\sigma_\mu)
\label{L_A-spin}
\end{eqnarray}
This is {\it identical} to ${\cal L}_{BCS}$
under the identification:  $\Phi = \Phi_\sigma$, $a_\mu = a_\mu^\sigma$
and $M^2= \mu^2 = r_\sigma$.
The upshot is that
a simple re-fermionization of the spin sector,
${\cal L}_\sigma$, gives directly the BCS quasiparticle
Lagrangian
${\cal L}_{BCS}$ in Eq.~\ref{eq:BCSham}.
The spin carrying but charge neutral vortex field, $\Phi_\sigma$,
is thus seen to be equivalent to a ``nodon" destruction operator.

By such a re-fermionization
procedure, we can infer the properties
of the vortex condensed phase, $\langle \Phi_\sigma \rangle \ne 0$,
with $r_\sigma$ negative.  This corresponds to taking
$\mu$ negative and entering the molecular 
limits of the $p_x + i p_y$ superconductor.
The
critical point at $r_\sigma =0$, with massless but uncondensed $\Phi_\sigma$,
is equivalent to the single massless Dirac
field (with $\mu=0$) centered at zero momentum.
{\it Without} recourse to re-fermionization,
vortex condensation $\langle \Phi_\sigma \rangle \ne 0$ 
directly implies
a mass for $a_\mu^\sigma$
and a vanishing spin Hall conductivity, $\sigma_{xy}^s =0$ - 
the correct value
for the $p_x + ip_y$ molecular superconductor.  This internal consistency
gives us some confidence in the more general
validity of the dual Ginzburg-Landau formulation.

An alternate route from the BCS to molecular limit 
is possible by implementing a {\it duality} transformation
on the bosonic theory ${\cal L}_\sigma$,
which interchanges the two phases.
This can be achieved 
by expressing the bosonic three-current 
for the conserved spin ($\Phi_\sigma^\dagger
\Phi_\sigma^{\vphantom\dagger} = \psi^\dagger \psi$)
as the curl of a gauge
field, $\alpha_\mu$, and integrating out $a_\mu$.  After shifting
$\alpha_\mu \rightarrow \alpha_\mu + A_\mu^\sigma$ one thereby obtains,
\begin{equation}
{\cal L}_{Dual} = {1 \over 2} |(\partial_\mu -i \alpha_\mu
-iA_\mu^\sigma)\phi|^2 
+ r_\phi |\phi|^2 + u |\phi|^4 + {\cal L}_{cs}(\alpha_\mu) .
\end{equation}
Here $\phi$ creates a vortex in the field $\Phi_\sigma$.
Notice that the dual theory has the same form
as ${\cal L}_{\sigma}$ in Eq.~\ref{L_A-spin}, except
for the absence of the Chern-Simons term in $A_\mu^\sigma$.
Under duality, the ``ordered"
phase with $r_\sigma < 0$ (and $\langle \Phi_\sigma \rangle \ne 0$) maps
into the ``disordered"
phase for $\phi$ with $r_\phi > 0$. 
In this phase (the molecular limit) the dual theory
correctly predicts
$\sigma_{xy}^s = 0$,  due to the absence
of the $A_\mu^\sigma$ Chern-Simons term.
In terms of the original Dirac field this
duality is a particle/hole transformation, $\psi \rightarrow \psi^\dagger$,
which changes the sign of the Dirac mass, $\mu \rightarrow -\mu$.  
The self-dual
point where both $\phi$ and $\Phi_\sigma$ are critical,
corresponds to the {\it massless} Dirac theory.

$ $From the $p_x+ip_y$ molecular superconducting phase
with $\sigma_{xy}^s =0$, it is possible
to also proliferate and condense the {\it charge} vortex:
$\langle \Phi_\rho \rangle \ne 0$, which describes
a fully spin-gapped crystalline phase of spinless charge $e$
``holons".    
\begin{figure}[htb]
\hspace{0.5in}\epsfxsize=3.25in\epsfbox{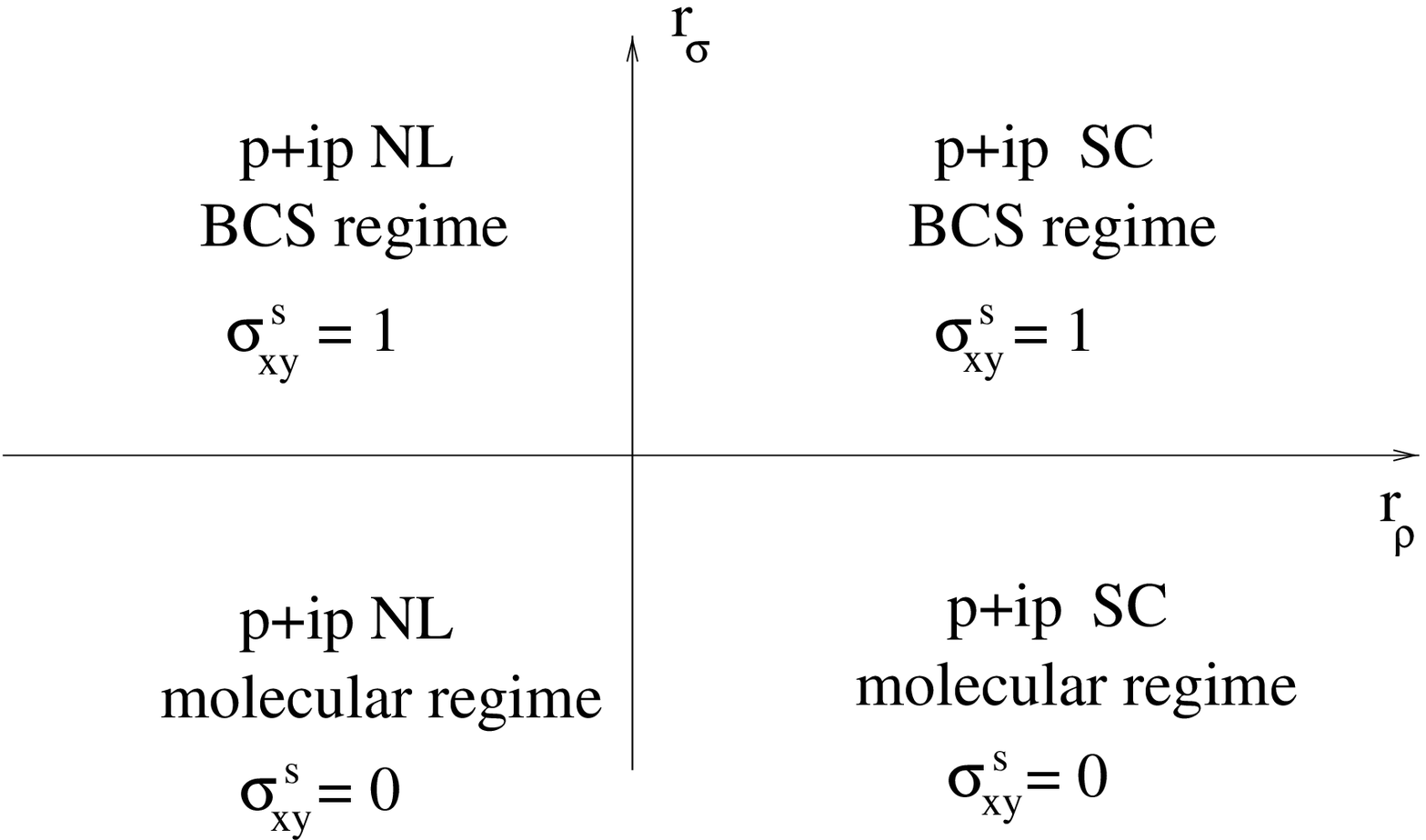}
\vskip 0.5 cm
{Fig.~1: The phase diagram in the ${r_\rho}-{r_\sigma}$
plane.  Here, nodal liquid is denoted
as $NL$, and $\sigma_{xy}^s$ is the spin
Hall conductivity. }
\end{figure}
In all four of the above phases (depicted schematically in the
$r_\rho$-$r_\sigma$ plane in Fig.~1)
time reversal invariance is spontaneously
broken, reflecting the underlying
$p_x +i p_y$ ``pairing" symmetry.
In each case, this symmetry breaking is
taking place in the {\it spin} sector of the theory.
The charge sector described by the
simple Ginzburg-Landau theory (with
{\it no} Chern-Simons term) is manifestly time
reversal invariant.  A natural question that arises
is whether the spin-charge separated dual vortex theory in
Eq.~\ref{eq:eff_Lag}\ 
can describe time reversal invariant phases,
such
as a $d_{x^2 - y^2}$ superconductor or nodal liquid.
Clearly a $d_{x^2 - y^2}$ phase requires
the breakdown of rotational invariance,
either spontaneously or by the presence of
an ionic potential.  Moreover, since the $d_{x^2 - y^2}$
superconductor exhibits gapless nodal excitations
with finite momentum (on the Fermi surface in weak coupling BCS),
it is necessary to access Fermi surface physics
at non-zero momenta in the dual formulation. 
This can be readily achieved as we now discuss.

\section{``Fermi surface", $p_x$, and $d_{x^2-y^2}$ phases}

\subsection{$p_x$ state}

In this section we imagine introducing an ionic potential with the
lattice symmetry.  The simplest case is a uniaxial perturbation, as
might be appropriate, e.g. in a quasi-one-dimensional superconductor
such as the cuprate ladder materials
(the more interesting generalization to square symmetry
will be returned to later).  For simplicity, we take the symmetry axes
along $x$ and $y$.  Physically, it is clear that such a potential
favors the formation of a {\sl real} (non T-breaking) paired state
such as $p_x$ or $p_y$.  This can be seen by considering the two-body
problem deep in the molecular limit.  In the presence of the lattice
potential, angular momentum is no longer a good quantum number (only
discrete $\pm \pi$ rotations and reflections are symmetry operations)
and the $p_x \pm i p_y$ states will generally be mixed.  Coupling the
two via a small ``tunneling'' perturbation, as appropriate for a {\sl
  weak} ionic potential, splits the two initially degenerate levels
into non-degenerate $p_x$ or $p_y$ eigenstates.  The system then
condenses into the lower of the two states.  In the BCS limit the
effects are more subtle, as we now illustrate.

As alluded to in the previous section, a distinguishing characteristic 
of the BCS theory of unconventional superconductors is the presence 
of gapless quasiparticle excitations at the intersections of the nodal 
lines of the pair wavefunction with the Fermi surface.  For the $p_x$
case, this occurs at two antipodal points in momentum space ${\bf k} = 
\pm K {\bf\hat{y}}$, and the resulting gapless quasiparticles can be
cast into the form of two Dirac species (see below).  The multiplicity
of Dirac fermions signals the emergence of {\sl two} conserved U(1)
charges in the low-energy limit.  Essentially, because of phase space
restrictions, the spin at each nodal point is separately conserved --
equivalently, one may view the two conserved charges as spin and
(quasi-)momentum, the latter obtaining due to the absence of
momentum-nonconserving umklapp processes for generically situated
nodal wavevectors.  

In the dual theory of Eq.~\ref{eq:eff_Lag}, total spin conservation
is manifest owing to the transverse nature of the spin current
$J^\sigma_\mu = {1 \over 2\pi}\epsilon_{\mu\nu\lambda} \partial_\nu
a^\sigma_\lambda$.  Momentum 
conservation is also clear, but apparently on a very different footing
-- it follows from the space-time Galilean invariance of the model.
Apparently, to describe the BCS limit, a connection must be made
between the internal (U(1) or SU(2)) symmetry of the model and the
external space-time translational symmetry.  It is quite remarkable
that such a connection can indeed be clarified, as we now show.

To see how the finite momentum physics can emerge in the dual theory,
first recall the quasiparticle structure of the $p_x+i\epsilon
p_y$ superconductor in the BCS limit.  This is described by
Eq.~\ref{eq:BCSham}\ with $\mu>0$ and $\Delta_{\bf k} = v_\Delta k_x + 
i \epsilon k_y$ (only the symmetry, and not the particular form of
this gap function is important in what follows).  The low energy
quasiparticles occur near $k_x=0$, $k_y = \pm K$, where $K \equiv
\sqrt{2m_e\mu}$.  To focus on these two regions in momentum space, next
define ``slowly-varying'' quasiparticle fields $f_{s\alpha}$ via
\begin{eqnarray}
  c_\uparrow({\bf x}) & \sim & \sum_{s=\pm} {1 \over \sqrt{2}}\left(
    f_{s\uparrow} - i f_{s\downarrow}\right)e^{is K y}, \label{eq:ddef1} \\
  c_\downarrow({\bf x}) & \sim & \sum_{s=\pm} {1 \over \sqrt{2}}\left(
    if_{-s\uparrow}^\dagger 
    + f_{-s\downarrow}^\dagger \right)e^{is K y}. \label{eq:ddef2}
\end{eqnarray}
One thereby obtains the Lagrange density
\begin{equation}
  {\cal L}_{p+i\epsilon p} = \sum_{s=\pm} f_s^\dagger
  \left[\partial_\tau + 
    iv_\Delta \tau^x \partial_x + is v_F \tau^y\partial_y + s m
    \tau^z\right] f_s^{\vphantom\dagger},
  \label{eq:2diracs}
\end{equation}
where $v_F = K/m_e$, $m=\epsilon K$ and, following
Eq.~\ref{BCS-Dirac}, we have 
introduced Pauli matrices $\vec{\tau}$ which act in the
$\alpha=\uparrow,\downarrow$ space.  Eq.~\ref{eq:2diracs}\ has the
form of two decoupled Dirac equations, and thereby displays
{\sl two} manifest U(1) symmetries.  Note from Eq.~\ref{eq:ddef2}\
that the $f_\downarrow$ fields are defined as hermitian conjugates
relative to the $f_\uparrow$ fields in Eq.~\ref{eq:ddef1}.  This
implies that the overall U(1) rotation, $f_s \rightarrow e^{i\chi}
f_s$, corresponds to spin ($S^z$) conservation.  The {\sl relative}
U(1), $f_s \rightarrow e^{is\chi} f_s$, embodies instead translational 
symmetry, or momentum conservation.  

The conserved densities are determined by N\"oethers theorem:
\begin{eqnarray}
  S^z_+ & = & f_+^\dagger f_+^{\vphantom\dagger} =
    c_{K\uparrow}^\dagger c_{K\uparrow}^{\vphantom\dagger} -
    c_{-K\downarrow}^\dagger c_{-K\downarrow}^{\vphantom\dagger} , \\
  S^z_- & = & f_-^\dagger f_-^{\vphantom\dagger} =
    c_{-K\uparrow}^\dagger c_{-K\uparrow}^{\vphantom\dagger} -
    c_{K\downarrow}^\dagger c_{K\downarrow}^{\vphantom\dagger}.
\end{eqnarray}
These are {\sl chiral} spin densities, closely analogous to the chiral 
densities encountered in one-dimensional Fermi systems.  The total
spin density $S^z = S^z_+ + S^z_-$.  

We are now in a position to bosonize the $p_x+i\epsilon p_y$ BCS
model.  The two flavors of Dirac particles $f_\pm$ may be traded for
two complex boson fields $\Phi_{\pm}$ by attaching flux in a variety of
ways.  The most natural, however, is to introduce a single gauge
field, attaching one flux quantum to each overall ($S^z$) U(1) charge.
This choice ensures both that all the resulting bosonic fields commute
at different space-time points {\sl and} that the gauge field has the
same physical meaning as the $a_\mu^\sigma$ defined previously.  In
particular,
\begin{equation}
  S^z = {1 \over {2\pi}} \bbox{\nabla} \times \bbox{a}^\sigma  .
\end{equation}
With this choice, the bosonized Lagrangian is
\begin{eqnarray}
  {\cal L}_{p+i\epsilon p} & = & \sum_s \left|D_0\Phi_{ s}\right|^2  +
  v_\Delta^2\left|D_1\Phi_{ s}\right|^2  +
  v_F^2\left|D_2\Phi_{ s}\right|^2
  \nonumber \\
  & & +
  W(\Phi_{ \pm}) + {\cal
    L}_{cs}(a^\sigma) - {\cal L}_{cs}(A^\sigma),
  \label{eq:2dbose}
\end{eqnarray}
where $D_\mu = \partial_\mu - i(a_\mu^\sigma + A_\mu^\sigma)$.
Here we have included an external spin gauge field $A^\sigma_\mu$ for
bookkeeping purposes.  The potential $W$ is dictated by symmetry to
take the form
\begin{eqnarray}
  W(\Phi_{ \pm}) & = & m^2\left[|\Phi_{ +}|^2 +
    |\Phi_{ -}|^2\right] \nonumber \\
  & & + u\left[|\Phi_{ +}|^2 +
    |\Phi_{ -}|^2\right]^2 +
  v|\Phi_{ +}\Phi_{ -}|^2 .
\end{eqnarray}
The final term in Eq.~\ref{eq:2dbose}\ is dictated by requiring
$\sigma_{xy}^s = 1$ in the $p_x+i\epsilon p_y$ phase where $m^2>0$.  

Remarkably, the selection of two such non-zero ``nodal'' points occurs
fairly naturally in our dual Ginzburg-Landau theory once uniaxial
(rectangular) anisotropy is included.  Consider the modified version of
Eq.~\ref{L_A-spin},
\begin{eqnarray}
  {\cal L}_\sigma & = & |D_0\Phi_\sigma|^2 + |D_1\Phi_\sigma|^2 +
  c|D_2\Phi_\sigma|^2 \nonumber \\ & & + {d \over 2}|{\bf D}^2\Phi_\sigma|^2 +
  r_\sigma|\Phi_\sigma|^2 + u_\sigma |\Phi_\sigma|^4 \nonumber \\
  & & + {\cal L}_{cs}(a^\sigma) - {\cal L}_{cs}(A^\sigma).
  \label{eq:uniaxial}
\end{eqnarray}
For simplicity, we have included only a single symmetry-breaking term, 
the coefficient $c<1$, which favors fluctuations along the $y$ axis over the
$x$ axis.  The coefficient $d>0$ is included for stability purposes.

As $c$ is decreased from one (zero anisotropy), the energy cost for
fluctuations of $\Phi_\sigma$ with spatial variations along $y$
becomes more and more reduced.  When $c$ changes sign and becomes
negative, the lowest energy fluctuations bifurcate away from the
origin in momentum space and move to two points ${\bf k} = \pm K
{\bf\hat{y}}$, with $K= \sqrt{|c|/d}$.  From this
point on, it is appropriate to focus on the low-energy field
configurations, viz
\begin{equation}
  \Phi_\sigma({\bf x}) \sim \Phi_{+} e^{iKy} +
  \Phi_{-}e^{-iKy}. 
  \label{eq:Phidecomp}
\end{equation} 
The physical meaning of Eq.~\ref{eq:Phidecomp}\ is clear from the
above ``reverse engineering'' of the field content of the $p_x + i
\epsilon p_y$ superconductor -- compare with
Eqs.~\ref{eq:ddef1}-\ref{eq:ddef2}.  Naively inserting
Eq.~\ref{eq:Phidecomp}\ in Eq.~\ref{eq:uniaxial}\ and neglecting
rapidly oscillating terms in the usual way gives an
effective Lagrangian for the $\Phi_{\sigma\pm}$ fields.  This has
{\it precisely} the form of Eq.~\ref{eq:2dbose}, with $v_\Delta^2 = 1+|c|$,
$v_F^2 = 2|c|$, $m^2= r_\sigma - c^2/2d$, $u=u_\sigma$, and
$v=2u_\sigma$.  These values (particularly $u$ and $v$) should,
however, not be taken too seriously, as they certainly depend upon the
simplistic treatment of fluctuations and higher-order terms.

With this identification in hand, we conclude that
Eq.~\ref{eq:uniaxial}\ provides a unified description of the $p_x \pm
ip_y$ and $p_x$ states in an intrinsically anisotropic system.  For
$c<0$, the equivalence to Eq.~\ref{eq:2dbose}\ allows a
refermionization to the form in Eq.~\ref{eq:2diracs}.  It is natural
to associate the critical point of these equations ($m=0$) with the
spin structure of the $p_x$ superconducting state and its
nodal-liquid/holon lattice counterpart.  The refermionized double
Dirac form in Eq.~\ref{eq:2diracs}\ is our primary result for the
uniaxially anisotropic model.

Issues of time-reversal symmetry merit some discussion.  Ideally, a
general formulation of the problem should contain T-non-invariant
terms only through spontaneous symmetry breaking.  However,
generically both Eq.~\ref{eq:2dbose}\ and Eq.~\ref{eq:2diracs}\ break
T.  In the fermionic formulation, Eq.~\ref{eq:2diracs}, fortunately,
the explicit symmetry breaking can be easily restored
by requiring $m^2=0$.
Furthermore, all T-preserving perturbations of this form can be 
shown to be irrelevant, so that the $p_x$ state is locally
stable.\cite{unpub}\ From the point of view of Eq.~\ref{eq:2dbose},
this is remarkable.  Indeed, a direct mean-field analysis would
suggest that a gapless state occurs {\sl only} along the critical line
$m^2=0$, requiring tuning of a parameter.  These conclusions can be
reconciled by noting that constraints upon Eq.~\ref{eq:2dbose}\ in a
T-invariant system are not at all obvious.  Only because of the
ability to refermionize are we able to identify the critical line
$m^2=0$ as containing a T-invariant manifold.  If time-reversal is
explicitly or spontaneously broken, a critical state is indeed
non-generic, and the spectrum of Eq.~\ref{eq:2dbose}\ correctly
reproduces that of the Dirac theory for small non-zero $m^2>0$
(i.e. $p_x+i\epsilon p_y$ with $\epsilon \ll 1$).
Similar subtleties render the analysis of Eq.~\ref{eq:2dbose}\ 
problematic for $m^2<0$.  We suspect that this region represents
rather more exotic T-violating states, and do not consider it further.

Another physical route away from the BCS-$p_x$ phase is via a
molecular $p_x$ state.  Indeed, as we have argued above, deep into the
molecular limit, uniaxial anisotropy guarantees a $p_x$ state.  To
tune through such a transition, we may imagine introducing a
finite-range attraction into the
`non-crossing' model which favors
tighter pair binding.  In such a model the $p_x$--BCS state naturally
undergoes a transition into a molecular $p_x$ state as the attraction
is increased.  From the conventional BCS point of view, we would
expect such a transition to be described by taking the chemical
potential through zero in a $p_x$ quasiparticle Hamiltonian, i.e.
Eq.~\ref{eq:BCSham}\ with $\Delta_k = v k_x$.  For this model, as
$\mu$ passes from positive to negative, the quasiparticle nodes
converge and coalesce at the origin, becoming massive for $\mu<0$.
Precisely at the critical point, one expects a spectrum $\omega^2
\approx v^2 k_x^2 + k_y^4/4m_e^2$.  This unconventional non-Lorentz
invariant form is precisely what is obtained at small wavevectors from
Eq.~\ref{eq:uniaxial}\ at the point where $c=r_\sigma=0$.  Just as the
gapless $p_x$ state occurs only along a line in mean-field theory but
comprises a phase in the gauge model, it appears that the critical
state at $c=r_\sigma=0$ (see Fig.~2)
in reality forms a phase boundary despite appearing as a
multicritical point in the mean-field treatment.

Given the existence of the $p_x$ molecular state, there must be
another critical line separating this from the $p_x \pm ip_y$
molecular phase.  As neither state contains gapless spin excitations
or possesses a non-zero $\sigma_{xy}^s$,
we have been unable to discriminate between them within the GL model.  Viewed 
from the point of view of self-consistent BCS theory, the transition
would appear to be a simple first-order level crossing of two-particle 
bound states.  Some preliminary modeling\cite{unpub}\ suggests that a
continuous transitions in the universality class of the quantum
transverse-field Ising model is also possible, the Ising order
parameter reflecting the presence or absence of T-breaking.  The full
proposed phase-diagram in the spin-sector is indicated in Fig.~2.

\begin{figure}[htb]
\hspace{0.5in}\epsfxsize=3.25in\epsfbox{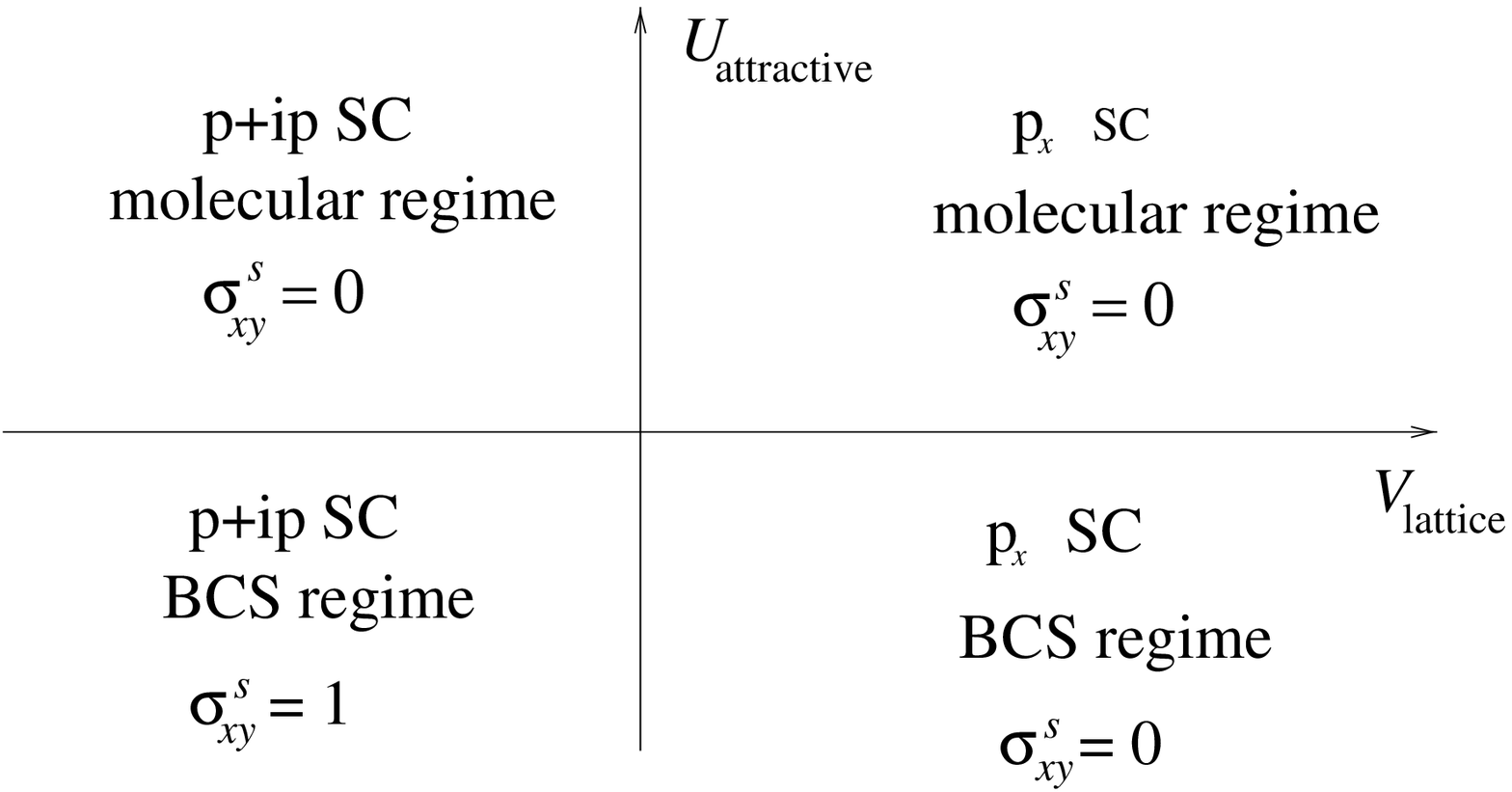}
\vskip 0.5 cm
{Fig.~2: The phase diagram as a function
of increasing lattice potential with uniaxial anisotropy, $V_{\rm lattice}$,
and short-range attractive pairing force, $U_{\rm attractive}$.}
\end{figure}

\subsection{square symmetry and $d$-wave}

We have seen how a BCS-like $p_x$ state with gapless modes at
non-zero momentum 
can emerge from the spin boson Lagrangian in Eq.~\ref{L_A-spin}\ in
the presence of uniaxial anisotropy.  If
the underlying crystal lattice has square symmetry, however, the $p_x$ 
or $p_y$ states are much less likely, and we expect instead that
$d$-wave pairing is favored.  Indeed, two particles in a fourfold
rotationally invariant lattice potential interacting with a hard core
will generally have a $d$-wave ground state in the limit of a strong
potential.  For a weak anisotropy, however, the $p_x \pm ip_y$ states are
likely lower in energy.  One thereby expects a transition upon
increasing lattice coupling between p-wave and d-wave pairing.  In the 
molecular limit this presumably occurs as a level crossing, i.e. a
first order phase transition.  In the BCS-like regime, however,
the nature of the transitions or sequence of transitions between these 
states is less clear.  

Rather than attempting to fully characterize this evolutionary
process, we will content ourselves instead with determining the spin
Lagrangian of the d-wave states themselves.  Following the logic of
the previous subsection, consider the simplest modification of
Eq.~\ref{L_A-spin}\ appropriate for square anisotropy,
\begin{eqnarray}
  {\cal L}_\sigma & = & |D_0\Phi_\sigma|^2 + c|D_i\Phi_\sigma|^2 +
  {d_1}|{\bf D}^2\Phi_\sigma|^2 \nonumber \\ & & + d_2
  \left(|D_1^2\Phi_\sigma|^2+ |D_2^2\Phi_\sigma|^2\right)  +
  r_\sigma|\Phi_\sigma|^2 + u_\sigma |\Phi_\sigma|^4 \nonumber \\
  & & + {\cal L}_{cs}(a^\sigma) - {\cal L}_{cs}(A^\sigma).
  \label{eq:square}
\end{eqnarray}
Here $d_2$ is a measure of the anisotropy, and low-energy excitations
are pushed to finite momentum if $c$ becomes negative.  For $c<0$ and
$d_2<0$, the lowest energy modes occur at four points ${\bf k} = (\pm
K,0), (0, \pm K)$, precisely as expected for the low-energy excitations in a
$d_{xy}$ superconductor (the momenta appropriate for $d_{x^2-y^2}$ are 
obtained for $d_2>0$)!  We fix $c<0$ and $d_2<0$ and consider
decreasing $r_\sigma$.  

Once again, it is appropriate to focus on the
four slowly-varying fields $\Phi_{ js}$ ($j=1,2$, $s=\pm$), defined by
\begin{equation}
  \Phi_{\sigma}({\bf x}) \sim \sum_{j s} \Phi_{ j s}e^{i s {\bf
      K}_j \cdot {\bf x}},
\end{equation}
where ${\bf K}_1 = (K,0)$, ${\bf K}_2 = (0, K)$, and
$K=\sqrt{|c|/(2(d_1+d_2))}$.  As for the p-wave case, an effective
theory can be developed for the $\Phi_{ j s}$ fields.  It takes
the form
\begin{eqnarray}
  {\cal L}_{eff} & = & \sum_{j s} \left|D_0\Phi_{ j s}\right|^2  +
  v_F^2\left|D_{j}\Phi_{ j s}\right|^2  +
  v_\Delta^2\left|\epsilon_{jj'}D_{j'}\Phi_{ j s}\right|^2
  \nonumber \\
  & & +
  \tilde{W}(\Phi_{1,2\pm}) + {\cal
    L}_{cs}(a^\sigma) - {\cal L}_{cs}(A^\sigma).
  \label{eq:dbose}
\end{eqnarray}
Neither the precise values of $v_F$, $v_\Delta$ nor the form of
$\tilde{W}$ is critical to this discussion.  What {\sl is}
significant, however, is the fact that Eq.~\ref{eq:dbose}\ takes the
form of four interacting relativistic complex bosons coupled to a
single U(1) Chern-Simons gauge field.  As before, the effect of this
gauge coupling is to attach identical flux to all spin quanta,
transmuting these bosons into four species of Dirac
fermions.  Furthermore, the spin Hall conductivity $\sigma_{xy}^s
=1$ in the massive phase (i.e. for $r_\sigma - c^2/(4(d_1+d_2)) >0$ in
mean-field theory), as determined by the last term in
Eq.~\ref{eq:dbose}.  This is in agreement with the value in the
$p_x\pm ip_y$ phase.  Here, this value gives some indication of the
structure of the {\sl signs} of the four Dirac masses in the
refermionized version of Eq.~\ref{eq:dbose}.  Each massive Dirac
equation gives a contribution of $\pm 1/2$ to $\sigma_{xy}^s$,
depending upon the sign of its mass term, so apparently there must be
three positive and one negative (or vice versa) masses in the Dirac
theory.  As the quadratic term in $\tilde{W}$ is tuned to zero, all
four bose (and hence Dirac) fields are expected to go critical.  This
is the natural candidate for a time-reversal invariant point, and we
speculate that it describes the nodal quasiparticles (nodons) in a
$d_{xy}$ superconductor and relatives such as the nodal
liquid.\cite{BFN1}\  The appropriate fermionic representation for the 
nodons in these states was derived independently in
Ref.~\onlinecite{BFN1}.  For $d_{xy}$ symmetry, it can be obtained as 
before from Eq.~\ref{eq:BCSham}\ by defining
\begin{eqnarray}
  c_\uparrow({\bf x}) & = & \sum_{j s} d_{j s
    \uparrow}^{\vphantom\dagger} e^{i s {\bf K}_j  
    \cdot {\bf x}}, \\
  c_\downarrow({\bf x}) & = & \sum_{js} d_{j s \downarrow}^\dagger
  e^{- i s {\bf K}_j \cdot {\bf x}}.
\end{eqnarray}
The (massless) nodon Lagrangian is then
\begin{eqnarray}
  {\cal L}_{nodon} & = & \sum_{s=\pm} d_{1s}^\dagger \left[
    \partial_\tau +  s v_F\tau^z i \partial_1 + s v_\Delta \tau^x i\partial_2
  \right]d_{1s}^{\vphantom\dagger} \nonumber \\
  & & + d_{2s}^\dagger \left[ \partial_\tau + 
    s v_F\tau^z i \partial_2 +s v_\Delta \tau^x \partial_1
  \right]d_{2s}^{\vphantom\dagger},
  \label{eq:nodonlag}
\end{eqnarray}
where as in earlier equations the $\vec{\tau}$ matrices act in the
$\alpha = \uparrow,\downarrow$ subspace.  For an alternate, explicitly 
SU(2) invariant formulation, see Ref.~\onlinecite{BFN1}.  
Eq.~\ref{eq:nodonlag}\ has the desired form of four  Dirac
equations (for $s=\pm$, $j=1,2$).  The identification of
Eq.~\ref{eq:nodonlag}\ is supported by the excellent correspondence between the
quantum numbers and momenta of the gapless modes of
Eq.~\ref{eq:dbose}\ and those of the nodons.  In either case, there
are four conserved U(1) currents, the charges (time components) of
which are chiral spin densities, i.e. spin densities for particles
with momenta along $\pm {\bf K}_1, \pm {\bf K}_2$.  Mass terms taking
Eq.~\ref{eq:nodonlag}\ away from criticality can also be added, and
take the form of $d_{is}^\dagger \tau^y d_{is}^{\vphantom\dagger}$
operators.  The explicitly break time-reversal invariance, so that as
for the $p_x$ phase, we expect the $d_{xy}$ (and analogously
$d_{x^2-y^2}$) state to be locally stable.  

Given the complexity of the arguments in this section, it seems
appropriate to summarize what has been learned.  We have studied how
momentum space structure emerges from the bosonic Ginzburg-Landau
theory of the spin sector.  In doing so, we have not assumed (as in
previous work on the nodal liquid) local superconductivity, but
proceeded instead on very general grounds.  Once the soft-modes of
$\Phi_\sigma$ move to non-zero momenta, any incipient critical points
can invariably be expressed in terms of multiple Dirac fields.  When
the associated Dirac masses vanish, a time-reversal invariant
lagrangian is possible, and two such theories were identified with the
$p_x$ and $d_{xy}$ nodal states.  Furthermore, uniaxial and square
lattice anisotropies were seen to favor appropriate critical states,
even in very naive treatments of the Landau theory.  These arguments
provide a partial derivation of Eq.~\ref{eq:nodonlag}\ for a generic
time-reversal invariant d-wave (superconducting {\sl or} nodal liquid)
state, contingent only on the original hard-core assumption used to
allow the Chern-Simons flux attachment.

\section{Discussion}

In this paper, we have suggested that spin-charge
separation is a generic consequence of strong
repulsion between electrons in two-dimensions.
We are driven inexorably to this conclusion
by the following logic. (1) We note that if there are
repulsive interactions which are strong enough to
prevent electron trajectories from intersecting,
then we may transmute the electrons into
hard-core bosons interacting with a Chern-Simons gauge
field which attaches flux to spin. (2) Since the
up- and down-spin boson currents are separately conserved,
they can be written as the curls of two auxiliary
gauge fields. (3) Using (1) and (2), we formulate
an equivalent dual theory which is of Ginzburg-Landau form.
The auxiliary gauge fields are minimally coupled to
vortex fields. (4) The Ginzburg-Landau theory contains a $Z_2$
symmetry which, if broken by vortex condensation,
leads to spin-charge confinement, translational symmetry-breaking,
and `conventional' ordered phases such
as the AF and CDW. (5) In order to study phases
with spin and charge physics at different scales and
without translational symmetry-breaking, we contemplate
$Z_2$-symmetric phases in which vortex pairs
condense. A phase diagram of spin-charge
separated states is the upshot of crossing the
rubicon fed by these five tributaries.

By eschewing a conventional momentum-space approach
which assumes a Fermi surface, we have constructed
an effective field theory which does not
fall under the rubric of Fermi liquid theory.
The basic excitations of our theory are {\it topological
solitons} in vortex condensates\cite{topological}.
They are in no sense adiabatically connected to the electron
and hole excitations of a free Fermi gas.
They are also rather different from the
`holon' and `spinon' concepts which are
introduced to solve the Gutzwiller constraint\cite{gaugetheories}.
These objects are strongly-coupled and do not
appear to be soliton-like in character.
The topological spinless charge $e$ excitation
in our Ginzburg-Landau theory (a ``holon") has an antiparticle
with opposite charge, and at low energies
can decouple from the neutral spin $1/2$ Fermionic
``nodon" excitation.  Moreover, the ``holon"
already exists as a finite
energy excitation within the Mott insulator - doping
is not required.

Contrasting the present work with
our earlier construction of the nodal liquid,
we see that the ``holon" field is identical to
the charged soliton of the nodal liquid but the nodon
-- which descended from an {\it assumed} quasiparticle
at the nodes of a $d_{{x^2}-{y^2}}$ superconductor --
is now a concept which naturally arises from the
spatially non-uniform softening of
a vortex-anti-vortex pair field.
Our nodon and holon fields are properly
seen as analogous to the charge and spin
solitons of the one-dimensional electron gas
or the fractionally-charge quasiparticles
in the quantum Hall effect.
 
Despite the surprising ease with which our dual Ginzburg-Landau
formulation captures spin-charge separation and superconductivity,
the Fermi liquid phase seems to be missing.  Generally, a dual 
vortex description 
of a Fermi liquid is possible, as illustrated nicely
for the case of {\it spinless} electrons.  After
transforming to spinless bosons via Chern-Simons
and implementing a duality transformation, one readily
obtains a simple dual Ginzburg-Landau theory.
This theory closely resembles Eq.~\ref{eqn:GL_Lag},
but with only a single vortex field, $\Phi$,
which is minimally
coupled to
a single gauge field with Chern-Simons dynamics.
But more importantly, the theory is non-relativistic (i.e. there is a
non-zero chemical potential) and the gauge field necessarily has a non-zero
average, as the dual flux equals the number of electrons.
So a vortex ``vacuum" phase - 
the $p_x + ip_y$ superconductor
for spinful electrons --  is not 
accessible without spin.  The Wigner crystal phase
of spinless electrons corresponds simply
to condensing the single vortex field, $\langle \Phi \rangle \ne 0$.
In the Fermi liquid phase this vortex field must
remain uncondensed, but with the vortices in a fluid state.  
This fluid of vortices presumably coexists with
the fluid of particles (the Chern-Simons
bosons which are the dual flux tubes) -- the particle
motion acting to scramble the vortex phase
and vice versa.

By analogy, a description of a {\it spinful} Fermi liquid via the dual
Ginzburg-Landau theory (Eq.~\ref{eqn:GL_Lag}) presumably requires an
uncondensed but ``critical" fluid of both up and down spin vortices.
But this analogy is problematic for two reasons: Firstly, both vortex
fields, $\Phi_\alpha$, are {\it relativistic}, with an equal number of
positive and negative circulation vortices.  But more importantly, in
Eq.~\ref{eqn:GL_Lag}\ the Chern-Simons term couples to the electron
{\it spin}, effectively mediating a long-range statistical interaction
between spin up and spin down electrons.  In a spinful Fermi liquid,
the up and down spin quasiparticle excitations should essentially
propagate independently.  It is for these reasons that we suspect
the impossibility of describing
a Fermi liquid phase in the dual formulation.

Why should the dual Ginzburg-Landau theory
preclude a Fermi liquid, given that the only explicit assumption
needed for its derivation was that opposite spin electron coordinates
never coincide?  In principle, the composite boson theory in
Eq.~\ref{eq:composite_boson}\ is exactly equivalent to a hard-core
interacting fermion model, and therefore has a Fermi liquid phase in
the presence of hard-core interactions {\sl only}.  The latter
conclusion can be understood from the two-body problem of a
small hard-core scatterer, from which it is seen that the s-wave
scattering amplitude is small when the hard-core radius $a \ll k_F$.
The subsequent manipulations leading to the dual formulation, however,
do not treat this constraint carefully.  Since the flux attachment
procedure itself adds relative angular momentum to the wavefunction
(c.f. Eq.~\ref{eq:CStransform}), to accomodate a significant s-wave
component requires incorporating {\sl p-wave} effects in the composite
bosons.  Such non-local physics is easily missed, and presumably
requires a much more subtle treatment of the duality transformation
condition to capture it. When the repulsive
interactions are sufficiently strong, however,
we suspect the dual lagrangian to be adequate.

As we have argued, the presence
of electrons spinning around one another is tantamount
to significant (finite angular momentum) pairing correlations.
The kinetic energy
clearly favors lower angular momentum,
which suggests the predominance of $p_x + ip_y$ pairing,
at least in the absence of significant ionic potentials.
Naively, this reasoning might suggest that
quantum Hall systems
with spinful electrons could
exhibit high-temperature superconductivity.
However, the presence of the
strong orbital
magnetic field presumably precludes this possibility.
But there are quantum Hall systems
which apparently do exhibit
high temperature ``pseudo-spin" superfluidity\cite{doublelayer}.
In particular, double-layer quantum Hall systems
with total filling $\nu=1$ do exhibit evidence
of a ``transverse superfluid" phase,
with superfluid currents in the two layers flowing
readily in {\it opposite} directions (effectively
negating the effects of the magnetic field).
It has been suggested that this phase will
disorder via a finite temperature
Kosterlitz-Thouless transition,
with $T_{KT}$ in the $1/2$ Kelvin range --
a superfluid
transition driven by Coulomb repulsion.

The rather close analogy
between
this quantum Hall system and the 2d $p_x + ip_y$ superconductor
(under the exchange of spin and charge) suggests a promising
route for studying {\sl energetics}.  In the quantum Hall effect,
variational wavefunctions have provided a powerful means of comparing
various candidate ground states.  Interchanging charge and spin, we
therefore propose the following candidate wavefunction for the
$p_x+ip_y$ superconductor:
\begin{equation}
  {\Psi}_{11\overline{1}} =
  {\prod_{i>j}}{\left({z_i} - {z_j}\right)}{\left({w_i} - {w_j}\right)}
  {\prod_{i,j}}{\left({{\overline z}_i} - {{\overline w}_j}\right)}
  {e^{- \gamma\sum {|z_i|^2} + {|w_i|^2}}}.
  \label{eq:111bar}
\end{equation}
Here, $z_i = x_i + i y_i$ denotes the complex coordinate
of an up spin electron, and $w_i$ are down spin electron coordinates.
The overbar denotes complex conjugation.
As for quantum Hall wavefunctions the electron density
is set by the exponential terms (with $\gamma$ proportional to
the 2d density).  But this wave function involves
{\it both} $z$ and it's complex conjugate $\overline{z}$.
In this way, electrons moving with large angular momentum 
are avoided.
In a slight abuse of quantum Hall notation, we refer to the
wavefunction in Eq.~\ref{eq:111bar}\ as a ``$(1,1,\overline{1})$''
state.  Readers familiar with
Chern-Simons theory will recognize that Eq.~\ref{eq:111bar}\
encapsulates the universal content (``K-matrix'') of the
effective field theory in Eq.~\ref{eq:LeffCS}, and is therefore a
faithful representation of the $p_x+ip_y$ state.  
To make the ``pairing" more explicit,
it is instructive to use the Cauchy identity
to re-express this wavefunction as
\begin{equation}
  {\Psi}_{11\overline{1}} =
  {\rm Det}\left(\frac{1}{{z_i} - {w_j}}\right)
  {\prod_{i,j}}{\left|{z_i} - {w_j}\right|^2}
  {e^{-\gamma \sum {|z_i|^2} + {|w_i|^2} }}.
  \label{eq:cauchy}
\end{equation}
Eq.~\ref{eq:cauchy}\  displays $\Psi_{11\overline{1}}$ as the product of a BCS
wavefunction (projected onto a state of definite electron number) with pair
wavefunction $1/z$ and a Jastrow factor.    
The Jastrow factor keeps up and down spin electrons apart
appropriate for repulsively interacting
2d electrons,
but the first term evidently encapsulates
pairing correlations adequate for superconductivity.
It will be
interesting to explore the energetics of $\Psi_{11\overline{1}}$ in
realistic models of repulsively interacting Fermions.

\begin{figure}[htb]
\hspace{0.5in}\epsfxsize=3.25in\epsfbox{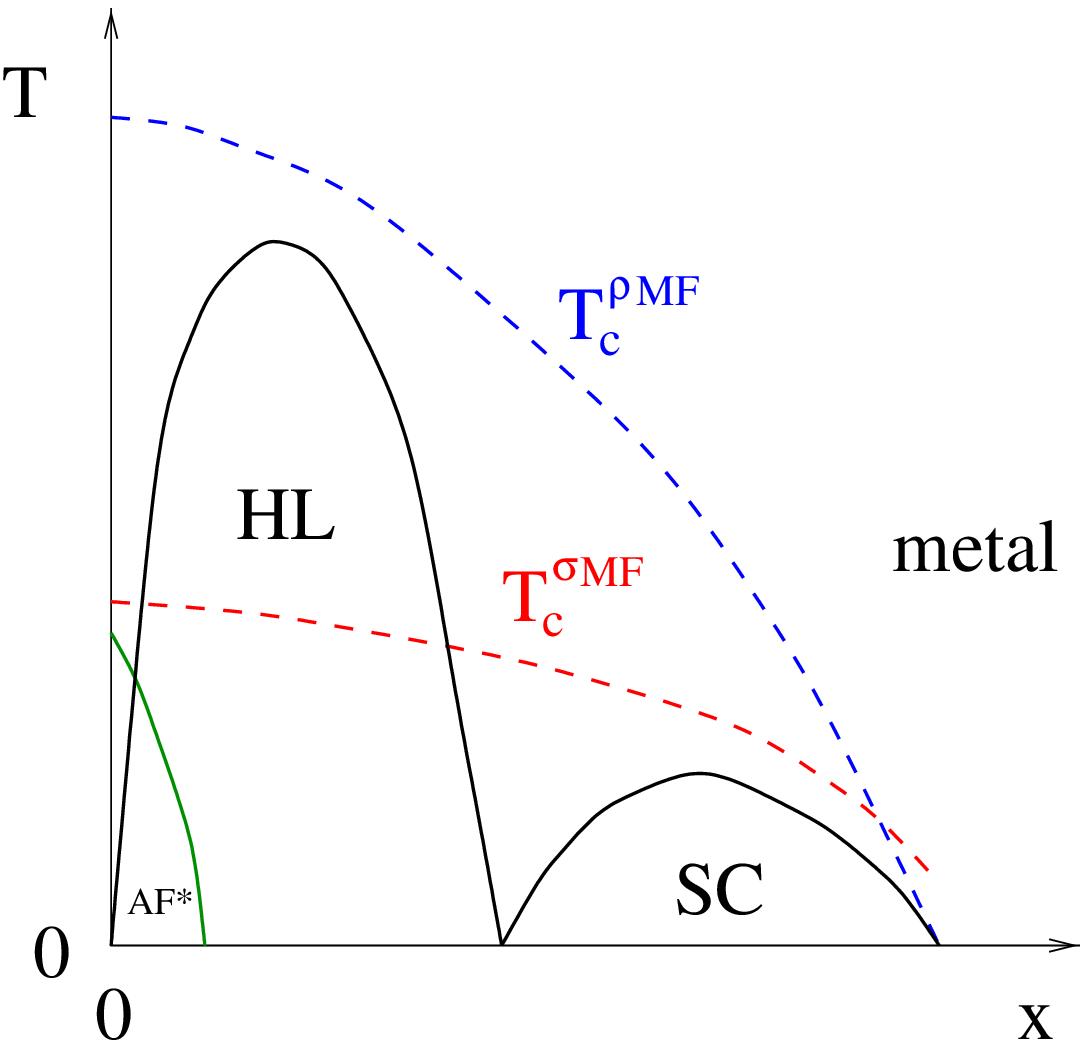}
\vskip 0.5 cm
{Fig.~3: Illustrative and schematic phase diagram of cuprate
  superconductors considered within the present dual Ginzburg-Landau
  framework.  The (unconventional) Antiferromagnet, Holon Lattice, and 
  SuperConducting phases are indicated by AF*, HL, and SC, respectively.}
\end{figure}

In the canonical approach to
strongly interacting electron systems near a Mott insulating
phase, the very first step is to
project onto a simpler tight binding
model, oftentimes with one orbital per unit cell.
When modeling the undoped cuprate superconductors
a further projection to
a reduced Hilbert space with one electron per
site is usually adopted.  The resulting spin Hamiltonian
is much more tractable than the full system of interacting electrons,
but we maintain that very important physics
is irretrievably lost under these projections.  
For instance, the spin-charge
separation that we access readily
within our dual Ginzburg-Landau formulation
of strongly interacting electrons
in the 2d {\it continuum} is certainly
not present in the Heisenberg spin Hamiltonian.
Put simply, it is exceedingly difficult to ascertain  
if spin and charge separate after projecting away
the charge.  We can, however, within our continuum approach,
still describe Mott insulating physics
by including a commensurate ionic potential
which locks the charge order.  In this way
it is possible to describe Mott insulators
which have gapped excitations and exhibit
spin-charge separation.  
Indeed much, if not all, of the interesting physics accessible
within the dual Ginzburg-Landau formulation
is inaccessible, and probably not present,
in the overworked $t-J$ model.

An apparent limitation of our approach is
that we do not have a specific microscopic model
which is described by our effective field theory
and we do not know {\it how strong} the
interactions must be to invalidate Fermi liquid theory.
Nevertheless, we expect that, since many of the
phases have excitation gaps, they should be stable to
small perturbations and there should be a universality
class of models which exhibit the same
``universal" and robust properties (such as
spin-charge separation).  Indeed, as emphasized previously, the only
{\sl formal} requirement for obtaining the dual formulation is the
``hard-core'' part of the repulsion.

The physics of the Ginzburg-Landau formulation shares some tantalizing
similarities with the cuprate high-$T_c$ materials.  As discussed
above, the crucial assumption of unbroken $Z_2$ symmetry appears most
natural for systems with a large native disparity between charge and
spin ordering scales.  This is indeed the case for the cuprates near
their half-filled Mott-insulating states.  Precisely at half-filling,
charge fluctuations begin to become quantized (acquire a gap) at very
high temperatures of order $eV$ (the Hubbard $U$), while for
spins the characteristic ordering energy scale is
significantly lower of order $J$
(with local moment formation occuring somewhat
higher).  As in our previous work,\cite{BFN1,BFN2}\  we hypothesize
that this Mott phase comprises a spin-charge separated insulator
described by Eq.~\ref{eq:eff_Lag}.  As
the electron density at half-filling is commensurate with the
underlying CuO$_2$ plane periodicity, the charge sector (as well as
the spin sector) is in effectively zero dual magnetic field.

The Ginzburg-Landau theories for charge and spin suggest a behavior in
the temperature-doping plane illustrated in Fig.~3.  At half-filling,
both vortex fields have zero external flux and make transitions from
their ``normal'' states at high temperatures to their ``Meissner''
states at low temperature.  The associated mean-field 
transition temperatures, roughly $T_c^{\rho MF}
\sim U$ and $T_c^{\sigma MF} \gtrsim J$, are
shown in Fig.~3.  Below $T_c^{\sigma MF}$, the spin
boson $\Phi_\sigma$ begins to develop amplitude fluctuations,
representing local moment formation.  At somewhat lower temperatures
this amplitude softens particularly near $(\pm \pi/2,\pm \pi/2)$, and
the refermionized Dirac fields subsequently order into an AF* phase
(see Ref.~\onlinecite{BFN2}\ for details).  Doping $x$ introduces a dual
external flux $\bbox{\nabla}\times \bbox{a}^\rho \propto \phi_0 x$ into the
charge sector only -- the spin boson $\Phi_\sigma$ is largely
unaffected and in particular $T_c^{\sigma MF}$ presumably decreases
only weakly.  The dual flux in the Ginzburg-Landau theory for
$\Phi_\rho$  introduces a dual mean-field ``$H_{c2}$''
line, or rapidly decreasing $T^{\rho MF}_c(x)$.  
Within a mean-field treatment
a holon lattice phase would be expected
below this line, in direct analogy
with the Abrikosov flux lattice.
But with fluctuations the holon lattice
phase should be separated from $T^{\rho MF}_c(x)$ by a
crossover regime analogous to the strongly-fluctuating ``vortex liquid"
state in type II superconductors.  In the cuprate context, this is a
regime of strong dynamical charge fluctuations and
can be thought of as a ``holon liquid", comprised of charge
$e$ bosons.  
The $T^{\rho MF}_c(x)$ line then represents a crossover
from a metallic phase above (with
``unquantized" charge) to the holon liquid which
manifests (dynamical) charge ``quantization"
in units of $e$ (cf. to dynamical ``flux quantization"
in the vortex liquid).  Remarkably,
our Ginzburg-Landau formulation
suggests that spin-charge separation,
effectively present below $T^{\rho MF}_c(x)$,
can occur on very high interaction energy scales
(eg. of order $U$). 
Upon further cooling the holon liquid one expects these bosons
to condense, provided their density
is sufficiently incommensurate with the
underlying crystal potential to avoid charge ordering
(into, for example, a holon lattice phase).
This is the superconducting state, expected
to be d-wave with a strong four-fold ionic potential.
The predominant
effect upon cooling through $T_c^{\sigma MF}$ above the superconducting
phase, should be a reduction
of low energy spin fluctuations and nodal formation
in the electron spectral function, with a lesser
effect in the charge sector due to
weak spin-charge couplings. 
Although we emphasize that this is very much a preliminary
application of the ideas of this paper, the picture in Fig.~3 is
suggestive. 

Our primary conclusion concerning
the ubiquity of spin-charge separation and
superconductivity driven by very strong repulsion has potential implications
for a much broader class of other strongly interacting systems.
Besides the cuprates, other
systems include: the heavy fermion superconductors\cite{heavy},
quasi-one dimensional organic superconductors, 
low carrier 2DEG's with very large $r_s$ in
semiconductor MOSFET's and heterostructures\cite{2d_mit,phillips},
superconductivity in  
$Sr_2RuO_4$ with possible $p_x + ip_y$ pairing
symmetry\cite{ruthenates},
the normal and superfluid phases of 3-He (the A phase
with a $p_x+ip_y$ pairing symmetry\cite{DMLee}) and perhaps most intriguingly
the magnetic states
of {\it solid} 3-He\cite{He3,DSFisher}.
In many of these systems one is also
very much interested in the full three dimensional
limit, particularly for 3-He.  
Unfortunately, the Chern-Simons
approach transforming fermions into bosons
by flux attachment is restricted to
strictly two-dimensional systems.  But
it is possible to transform
between fermions and bosons in three
dimensions by binding ``statistical" magnetic
monopoles to the particles\cite{Dyon}.
Unfortunately, this introduces an unphysical
internal statistical magnetic
field (in contrast to the pure gauge
coupling within Chern-Simons theory).  
But by attaching monopoles to {\it spin},
the monopole fields from the up spin electrons and the
antimonopole fields from the down spin electrons
will largely cancel (exactly on the average).  
Moreover, particle-vortex duality transformations
are also possible in three-dimensions (ie.
``electric-magnetic" duality), so it should
be possible to obtain an entirely bosonic (but approximate)
dual description of 3d electrons with a
`non-crossing' constraint.  Perhaps this approach might be
useful in modeling some 3d strongly correlated systems.

If, as we have suggested, strongly-interacting
spin-$1/2$ fermions
do not form a Fermi liquid,
then our effective field theory represents a new paradigm for correlated
electron behavior. If, as we have further
hypothesized, superconductivity is a prevalent
attribute of the phases ensconced in our theory, then
it is a paradigm which includes a new route to superconductivity.
For these reasons, we submit that that our scenario could
have far-reaching implications for the cuprate
superconductors and other strongly-interacting
electron systems.

We are grateful to Steve Girvin, T. Senthil and Doug Scalapino
for illuminating discussions, and Mohit Randeria for gentle criticism
of an overly hurried first draft.  
This research was generously supported by the NSF 
under Grants DMR-97-04005,
DMR95-28578
and PHY94-07194.

\end{document}